\documentclass[10pt,journal]{IEEEtran}

\usepackage{graphicx}
\usepackage{algorithmic}
\usepackage{algorithm}
\usepackage{amsgen}
\usepackage{amsfonts}
\usepackage{amsbsy}
\usepackage{amssymb}
\usepackage{amsmath}

\newcommand{\beq}{\begin{equation}}
\newcommand{\eeq}{\end{equation}}
\newcommand{\beqa}{\begin{eqnarray}}
\newcommand{\eeqa}{\end{eqnarray}}
\newcommand{\beqan}{\begin{eqnarray*}}
\newcommand{\eeqan}{\end{eqnarray*}}

\def\BibTeX{{\rm B\kern-.05em{\sc i\kern-.025em b}\kern-.08emT\kern-.1667em\lower.7ex\hbox{E}\kern-.125emX}}

\begin{document}

\title{Entropy-optimal Generalized Token Bucket Regulator}

\author{Ashutosh Deepak Gore and Abhay Karandikar\\
Information Networks Laboratory\\
Department of Electrical Engineering\\
Indian Institute of Technology - Bombay\\
\{adgore,karandi\}@ee.iitb.ac.in
}

\maketitle

\newcommand{\labelsection}{Section \Roman{section})}
\begin{abstract}
We derive the maximum entropy of a flow (information utility) which
conforms to traffic constraints imposed by a generalized token bucket
regulator, by taking into account the covert information present in
the randomness of packet lengths. Under equality constraints of
aggregate tokens and aggregate bucket depth, a generalized token
bucket regulator can achieve higher information utility than a
standard token bucket regulator. The optimal generalized token bucket
regulator has a near-uniform bucket depth sequence and a decreasing
token increment sequence.
\end{abstract}

\begin{keywords}
network information theory, token bucket traffic regulation, packet
length schedule, quality of service
\end{keywords}

\section{Introduction}
\label{intro}
In Internet Quality of Service (QoS) parlance, as a part of the
service level agreement (SLA) between a subscriber (source) and an
Internet service provider (ISP), a token bucket regulator (TBR) can be
used to smoothen the bursty nature of a subscriber's traffic
\cite{nw_keshav}.  The SLA mandates that the ISP provide end-to-end
loss and delay guarantees to a subscriber's packets, provided the
traffic profile of the subscriber adheres to certain TBR
constraints. The standard token bucket regulator (STBR), as defined by
the Internet Engineering Task Force (IETF), enforces
linear-boundedness on the flow and is characterized by the token
increment rate $r$ and the bucket depth $B$.  We will be more general
and consider a TBR in which the token increment rate and bucket depth
(maximum burst size) can vary from slot to slot. Such a TBR, which we
define as a generalized token bucket regulator (GTBR), can be used to
regulate variable bit rate (VBR) traffic\footnote{For example, a
pre-recorded video stream.} from a source \cite{ieeecl_premal}.  The
continuous-time analogue of a GTBR is the time-varying leaky bucket
shaper \cite{giordano_renegotiable} in which the token rate and bucket
depth parameters can change at specified time instants. In
\cite{giordano_renegotiable}, the authors determine the optimal
parameters (rates and bucket sizes) and apply it to the renegotiable
VBR service.

Our primary contribution is developing the notion of information
utility of a GTBR. Specifically, we derive the maximum information
that a GTBR-conforming traffic flow can convey in a finite time
interval, by taking into account the additional information present in
the randomness of packet lengths.  The idea of using a covert channel
to convey side information\footnote{Information present in packets
other than the actual packet contents.} in data networks has been
investigated earlier in the classic papers \cite{protocol_gallager}
\cite{ieeeit_verdu}. In this paper, the side information is considered
in the lengths of the packets only. Of all the packet length schedules
that conform to a given GTBR, our objective is to stochastically
characterize the flow that has the maximum entropy.

In \cite{iee_premal}, the authors have derived the information utility
of an STBR and suggested a pricing viewpoint for its application. Our
interest is more theoretical -- we consider an STBR as a special case
of a GTBR and describe a framework for their information-theoretic
comparison. We investigate whether a GTBR can achieve higher flow
entropy than an STBR and explain the properties of entropy-maximizing
GTBRs.

Section \ref{sysmod} explains our system model. In Section
\ref{iter-method}, we derive the optimal flow entropy equation and
define the information utility of a GTBR. In Section \ref{prb-frmltn},
we formulate the optimal GTBR and derive a necessary condition.  In
Section \ref{simulation1}, we compute the optimal GTBR. We interpret
our results in Section \ref{mnbvc}, and conclude in Section
\ref{dksjl}.

\section{System model}
\label{sysmod}

\begin{figure}[thbp]
{\centering \resizebox*{2in}{!}{\includegraphics{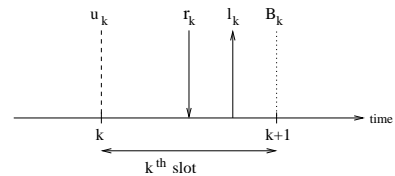}} \par}
\caption{Relative time instants of  parameters defined in (\ref{gowqk}).}
\label{slot-diagram}
\end{figure}

Consider a system in which time is divided into slots and a source
which has to complete its data transmission within $N$ slots. In our
discrete-time model, we will evaluate the system at time instants
$0,1,\ldots,N-1,N$. The $k^{th}$ slot is defined to be the time
interval $[k,k+1)$. The traffic from the source is regulated by a
GTBR. Define
\begin{eqnarray}
r_k &:=& \mbox{token increment for the $k^{th}$ slot} \nonumber \\
B_k &:=& \mbox{bucket depth for the $(k+1)^{th}$ slot} \nonumber \\
\ell_k &:=& \mbox{length of packet transmitted in the $k^{th}$ slot}\nonumber\\
u_k &:=& \mbox{residual  tokens at  start of the $k^{th}$ slot}
\label{gowqk}
\end{eqnarray}
$r_k$, $B_k$, $\ell_k$ and $u_k$, whose relative time instants are
shown in Figure \ref{slot-diagram}, are all non-negative
integers.  Let ${\mathbf r} := (r_0,r_1,\ldots,r_{N-1})$
denote the token increment sequence and ${\mathbf B} :=
(B_0,B_1,\ldots,B_{N-2})$ denote the bucket depth sequence.  The
system starts with zero tokens; $u_0=0$.  A GTBR $\mathcal R$
with the above parameters, written as ${\mathcal R}(N,{\mathbf
r},{\mathbf B})$, constrains the packet lengths according to
\begin{eqnarray}
\ell_i \leq u_i + r_i \; \; \; \forall \; i: 0 \leq i \leq N-1
\label{tok-ineq}
\end{eqnarray}
If (\ref{tok-ineq}) is satisfied, then ${\mathbf
\ell}=(\ell_0,\ell_1,\ldots,\ell_{N-1})$ is a conforming packet length
vector and $u_i$ evolves as
\begin{eqnarray}
u_{i+1} &=& \min(u_i + r_i - \ell_i, B_i) \; \; \forall \; 
    i: 0 \leq i \leq N-2 \nonumber \\
u_N &=& u_{N-1}+r_{N-1}-\ell_{N-1}
\label{tok-evlv}
\end{eqnarray}
If $r_i=r$ and $B_i=B$ for all $i$, then the GTBR ${\mathcal
R}_g(N,{\mathbf r},{\mathbf B})$ degenerates to the STBR ${\mathcal
R}_s(N,r,B)$.

\section{Information utility}
\label{iter-method}

Consider a source which has a large amount of data to send and whose
traffic is regulated by a GTBR. We seek to maximize the information
that the source can convey in the given time interval or the entropy
present in the source traffic flow in an information-theoretic
sense. The maximum entropy achievable by any flow which is constrained
by the GTBR ${\mathcal R}(N,{\mathbf r},{\mathbf B})$ is defined to be
its information utility.  The source can send information to the
destination via two channels:

\begin{enumerate}
\renewcommand{\theenumi}{\roman{enumi}}

\item
\label{overt} 
Overt channel: The contents of each packet.  Let $\ell_i$ be the
length of a packet in bits.  The value of each bit is $0$ or $1$ with
equal probability and is  independent of the values taken
by the preceding and succeeding bits.  The packet thus contributes
$\ell_i$ bits of information.

\item
\label{covert}
Covert channel: We consider the length of a packet as an event
and associate a probability with it.  Thus, side information is
transmitted by the randomness in the packet lengths.

\end{enumerate}
At time $k$, the only method by which past transmissions can constrain
the rest of the flow is by the residual number of tokens $u_k$.  The
key observation is that the future entropy depends only on the buffer
level $u_k$ at time $k$.  So, $u_k$ captures the state of the system.
Entropy is a function of system state $u_k$ and is denoted by
$H_k(u_k)$.

At time $N$, the source signals the termination of the current flow by
transmitting a special string of bits (flag).  The information
transmitted by this fixed sequence of bits is zero.
\begin{eqnarray}
\therefore H_N(u_N) &=& 0
\label{entropy-N}
\end{eqnarray}

\noindent For a given state $u_k$ of the system, if a packet of length $\ell_k$
bits is transmitted with probability $p_{\ell_k}(u_k)$, then:
\begin{enumerate}
\item The overt information transmitted is $\ell_k$ bits.
\item As the event occurs with probability $p_{\ell_k}(u_k)$,
the covert information transmitted is $(-\log_2 p_{\ell_k}(u_k))$ bits.
\item Since $\ell_k$ is random, $u_{k+1}$ is also random (from (\ref{tok-evlv})). 
Thus, $H_{k+1}(u_{k+1})$ is also a random variable.
\end{enumerate}
Adding all of the above and averaging it over all conforming packet
lengths, we obtain the entropy of the current stage:
\begin{multline}
H_k(u_k) = \sum_{\ell_k=0}^{u_k+r_k} p_{\ell_k}(u_k) \Big( \ell_k 
- \log_2(p_{\ell_k}(u_k)) \; + \\ H_{k+1}(\min(u_k+r_k-\ell_k,B_k)) \Big)
\; \; \forall \; k=0,\ldots,N-1 
\label{entropy-k}
\end{multline}
Finally, the above probabilities must satisfy
\begin{eqnarray}
\sum_{\ell_k=0}^{u_k+r_k} p_{\ell_k}(u_k) = 1 \; \; \forall \; k=0,\ldots,N-1 
\label{probsum-k}
\end{eqnarray}
Let ${\mathbf p}_k(u_k) :=
\left(p_0(u_k),p_1(u_k),\cdots,p_{u_k+r_k}(u_k)\right)$.  Our
objective is to determine the sequence of probability mass
functions\footnote{The dependence of $p_{\ell_k}$ and ${\mathbf p}_k$
on $u_k$ is assumed to be understood and is not always stated
explicitly. So, ${\mathbf p}_k =
\left(p_0,p_1,\cdots,p_{u_k+r_k}\right)$.}  $({\mathbf
p}_{N-1}^*,{\mathbf p}_{N-2}^*,\cdots,{\mathbf p}_0^*)$ which
maximizes the flow entropy $H_0(0)$ for a given GTBR ${\mathcal
R}(N,{\mathbf r},{\mathbf B})$. 
From (\ref{entropy-N})
\begin{eqnarray*}
H_N^*(u_N) &=& 0
\end{eqnarray*}
From (\ref{entropy-k})
\begin{multline*}
H_k(u_k) = \sum_{\ell_k=0}^{u_k+r_k} p_{\ell_k} \Big( \ell_k - 
  \log_2(p_{\ell_k}) + H_{k+1}^*(\min(u_k+\\
 r_k-\ell_k,B_k)) \Big) \; \; \forall \; k = 0, \ldots, N-1
\end{multline*}
Given $H_{k+1}^*(u_{k+1})$ $\forall$ $u_{k+1}$, there exists an
optimum probability vector ${\mathbf p}_k^* =
(p_0^*,p_1^*,\ldots,p_{u_k+r_k}^*)$ which maximizes the flow entropy
$H_k(u_k)$.
\begin{multline}
\therefore H_k^*(u_k) = \sum_{\ell_k=0}^{u_k+r_k} p_{\ell_k}^* \Big( \ell_k - 
  \log_2(p_{\ell_k}^*) + H_{k+1}^*(\min(u_k+\\
 r_k-\ell_k,B_k)) \Big) \; \; \forall \; k=0, \ldots, N-1 
\label{uwxphis}
\end{multline}
Thus, the problem of computing the entire sequence of probability
vectors $({\mathbf p}_{N-1}^*,{\mathbf p}_{N-2}^*,\cdots,{\mathbf
p}_0^*)$ has now been decoupled into a sequence of subproblems. 
The subproblem for time $k$ is:\\ Given the function
$H_{k+1}^*(u_{k+1}) \; \forall \; u_{k+1}$, determine the probability
vector ${\mathbf p}_k = (p_0,p_1,\ldots,p_{u_k+r_k})$ so as to
\begin{multline}
\mbox{maximize} \sum_{\ell_k=0}^{u_k+r_k} p_{\ell_k} \Big( \ell_k - 
  \log_2(p_{\ell_k}) + H_{k+1}^*(\min(u_k+r_k\\-\ell_k,B_k)) \Big)
\; \; \mbox{subject to} \; \; \sum_{\ell_k=0}^{u_k+r_k} p_{\ell_k}  = 1 
\label{opti-stage}
\end{multline}
(\ref{opti-stage}) can be solved using Lagrange multipliers.
\begin{multline}
{\mathcal L}({\mathbf p}_k,\lambda_k) := \sum_{\ell_k=0}^{u_k+r_k} p_{\ell_k} 
\Big( \ell_k - \log_2(p_{\ell_k}) + H_{k+1}^*(\min(u_k+r_k\\-\ell_k,B_k)) \Big)
+ \lambda_k \Big( \sum_{\ell_k=0}^{u_k+r_k} p_{\ell_k} - 1 \Big)
\end{multline}
At the optimal point $({\mathbf p}_k^*,\lambda_k^*)$
\begin{eqnarray}
\left. \frac{\partial{\mathcal L}}{\partial p_{\ell_k}} \right| 
\begin{array}{c} \\ ({\mathbf p}_k^*,\lambda_k^*) \end{array}
&=& 0 \; \; \; \;
\forall \; \ell_k = 0, \ldots, u_k+r_k \label{lagrange1} \\
\left. \frac{\partial{\mathcal L}}{\partial \lambda_k} \right| 
\begin{array}{c} \\ ({\mathbf p}_k^*,\lambda_k^*) \end{array} &=& 0
\label{lagrange2}
\end{eqnarray}
Solving (\ref{lagrange2})
\begin{eqnarray}
\sum_{\ell_k=0}^{u_k+r_k} p_{\ell_k}^*(u_k) = 1
\label{optprobunity}
\end{eqnarray}
Solving (\ref{lagrange1})
\begin{multline}
p_{\ell_k}^*(u_k) = 2^{\ell_k - \log_2 e + H_{k+1}^*(\min(u_k+r_k-\ell_k,B_k))
+\lambda_k^*(u_k)}
\label{opti-prb}
\end{multline}
From (\ref{optprobunity}) and (\ref{opti-prb})
\begin{multline}
\lambda_k^*(u_k) = \log_2 \bigg( \frac{e}{\sum_{\ell_k=0}^{u_k+r_k} 
2^{\ell_k + H_{k+1}^*(\min(u_k+r_k-\ell_k,B_k))}} \bigg)
\label{optml-lambda}
\end{multline}
From (\ref{opti-prb}) and (\ref{optml-lambda})
\begin{multline}
p_{\ell_k}^*(u_k) = 
\frac{2^{\ell_k + H_{k+1}^*(\min(u_k+r_k-\ell_k,B_k))}}
{\sum_{\alpha_k=0}^{u_k+r_k} 2^{\alpha_k + H_{k+1}^*(\min(u_k+r_k-\alpha_k,B_k))}}
\label{opti-prb2}
\end{multline}
From (\ref{uwxphis}) and (\ref{opti-prb2}), we finally obtain
\begin{multline}
H_k^*(u_k) = \log_2 \Big(
\sum_{\ell_k=0}^{u_k+r_k} 2^{\ell_k + H_{k+1}^*(\min(u_k+r_k-\ell_k,B_k))} \Big) 
\label{hkopti}
\end{multline}
Starting with $H_N^*(u_N)=0$, we use (\ref{hkopti}) to compute the
optimal flow entropy $H_k^*(u_k)$ for all $u_k$ and then proceed
backward recursively for $k=N-1,N-2,\ldots,0$.  The information
utility of the GTBR is $H_0^*(0)$.  

\section{Problem Formulation}
\label{prb-frmltn}
For the information-theoretic comparison of a GTBR ${\mathcal
R}_g(N,{\mathbf r},{\mathbf B})$ and an STBR ${\mathcal R}_s(N',r,B)$,
we impose the following conditions:

\begin{enumerate}
\renewcommand{\theenumi}{\alph{enumi}}

\item ${\mathcal R}_g$ and ${\mathcal R}_s$ must operate over
the same number of slots.
\begin{eqnarray*}
N=N'
\end{eqnarray*}

\item The aggregate tokens of ${\mathcal R}_g$ and ${\mathcal R}_s$
must be equal.
\begin{eqnarray}
\sum_{i=0}^{N-1} r_i = Nr \label{fiejs}
\end{eqnarray}

\item The aggregate bucket depth of ${\mathcal R}_g$ must not exceed
that of ${\mathcal R}_s$\footnote{Equality is present in (\ref{fiejs})
because every additional token directly translates to the permission
to transmit one more bit, leading to increase in information utility.
As this may not be necessarily true for bucket depth, we permit inequality
in (\ref{ukhef}).}.
\begin{eqnarray}
\sum_{i=0}^{N-2} {B_i} \leq (N-1)B \label{ukhef}
\end{eqnarray}

\item
The bucket depth of ${\mathcal R}_s$ cannot be very high compared to
its token increment rate.
\begin{eqnarray}
2r \leq B \leq 5r \label{sdidk}
\end{eqnarray}
For example, in \cite{giordano_renegotiable}, the authors use
$r_{max}=6 \mbox{ Mbps}$ and $B_{max}=12 \mbox{ Mbps}$ for their
simulations.

\item
The  token increment rate of ${\mathcal R}_g$ at every stage
must not be higher than the bucket depth of ${\mathcal R}_s$.
\begin{eqnarray}
r_i \leq B \label{minmi}
\end{eqnarray}

\end{enumerate}
\noindent The optimal GTBR problem is:\\
Given an STBR ${\mathcal R}_s(N,r,B)$, determine $\mathbf r$
and $\mathbf B$ of a GTBR ${\mathcal R}_g(N,{\mathbf r},{\mathbf B})$
so as to maximize $H_0^*(0)$ subject to (\ref{fiejs}), (\ref{ukhef}), 
(\ref{sdidk}) and (\ref{minmi}).\\
\newline \noindent The following result significantly reduces the search space for
the optimal GTBR.\\
\newline
\noindent {\bf Proposition:} For an optimal GTBR, equality must hold in  
(\ref{ukhef}), except when $N$ is small.\\
\newline
\noindent {\bf Proof:} We prove by contradiction.  Define
$g_k(u)=2^{H_k^*(u)}$.  Since $H_k^*(u) \geq 0$, $g_k(u) \geq 1$.  From
(\ref{hkopti}),
\begin{eqnarray}
g_k(u) &=& \sum_{\ell=0}^{u+r_k} 2^\ell g_{k+1}(\min(u+r_k-\ell,B_k)) \label{gfuwx}
\end{eqnarray}
$g_{N-1}(u)=2^{u+r_{N-1}+1}-1$ is an increasing sequence in $u$. Using
(\ref{gfuwx}), we can show that $g_k(u)$ is an increasing sequence in
$u$ $\forall$ $k=0,\ldots,N-1$.
Let $\phi_i=$ maximum number of tokens possible at time  $i$. Thus,
$\phi_0 = 0$ and
\begin{multline}
\phi_i = \min(\phi_{i-1}+r_{i-1},B_{i-1}) \; \forall i=1,\ldots,N-1
\end{multline}
If $u_i \leq \phi_i$, then we say that  state $u_i$ is 
reachable at stage $i$, otherwise it is unreachable.  

Let
${\mathcal R}(N,{\mathbf r},{\mathbf B})$ be an optimal GTBR, for
which equality does not hold in (\ref{ukhef}). Then
$\sum_{i=0}^{N-2}B_i \leq (N-1)B-1$. Consider another GTBR ${\mathcal
R}'(N,{\mathbf r'},{\mathbf B'})$ with ${\mathbf r'}={\mathbf r}$ and
${\mathbf B'}=(B_0,\ldots,B_{k-1},B_k+1,B_{k+1},\ldots,B_{N-2})$
for some $k$.
${\mathbf B'}$ satisfies (\ref{ukhef}).
$g_i'(u)=g_i(u)$ $\forall$ $i=k+1,\ldots,N$ and $\forall$ $u$.  Since
$\min(u+r_k-\ell,B_k+1) \geq \min(u+r_k-\ell,B_k)$,
$g_k(\min(u+r_k-\ell,B_k+1)) \geq g_k(\min(u+r_k-\ell,B_k)) \geq 1$.
If we determine a reachable state $u$ such that $g_k'(u) >
g_k(u)$, then $g_0'(0) > g_0(0)$, since the flow entropy at stage $0$
is computed stage-by-stage as a linear sum of future possible flow
entropies with positive weights.  Thus, the problem now reduces to
determining a stage $k$ and a reachable state $u$ such that $g_k'(u) >
g_k(u)$.  One of the following must hold:
\begin{description}
\item[{\it Case 1}] $\;$ There exists an $i$ $\in$ $\{1,\ldots,N-1\}$ 
 such that $\phi_i=B_{i-1}<\phi_{i-1}+r_{i-1}$.
\item[{\it Case 2}] $\;$ There is no  $i$ such that $\phi_i=B_{i-1}<\phi_{i-1}+r_{i-1}$.
\end{description}

\noindent {\it Case 1:} Consider the smallest $i$ such that 
$\phi_i=B_{i-1}<\phi_{i-1}+r_{i-1}$. Take $k=i-1$.
From (\ref{gfuwx})
\begin{multline}
g_{i-1}(u) = \sum_{\ell=0}^{u+r_{i-1}} 2^\ell g_i(\min(u+r_{i-1}-\ell,B_{i-1}))\\
 = \sum_{\ell=0}^{\stackrel{u+r_{i-1}}{-B_{i-1}-1}} 2^\ell g_i(B_{i-1})
  + \sum_{\stackrel{\ell=u+r_{i-1}}{-B_{i-1}}}^{u+r_i} 2^\ell g_i(u+r_{i-1}-\ell) \label{slajs}
\end{multline}
\begin{multline}
g_{i-1}'(u) = \sum_{\ell=0}^{u+r_{i-1}} 2^\ell g_i(\min(u+r_{i-1}-\ell,B_{i-1}+1)) =\\
\sum_{\ell=0}^{\stackrel{u+r_{i-1}}{-B_{i-1}-1}} 2^\ell g_i(B_{i-1}+1)
  + \sum_{\stackrel{\ell=u+r_{i-1}}{-B_{i-1}}}^{u+r_i} 2^\ell g_i(u+r_{i-1}-\ell) \label{ghgfu}
\end{multline}
(\ref{slajs}) and (\ref{ghgfu}) hold only if
\begin{eqnarray}
u+r_{i-1}-B_{i-1}-1 \geq 0
\label{ireewn}
\end{eqnarray}
$u=\phi_{i-1}$ is a state which is reachable in the original system 
as well as in the primed system and satisfies (\ref{ireewn}). 
Since $g_i(u)$ is an increasing sequence in $u$,  (\ref{slajs}) 
and (\ref{ghgfu}) imply
$g_{i-1}'(\phi_{i-1}) > g_{i-1}(\phi_{i-1})$. Consequently, $g_0'(0)>g_0(0)$.

\noindent {\it Case 2:} If no such $i$ exists, then
$B_i \geq r_0 + \cdots + r_i$ $\forall$ $i=0,\ldots,N-2$.
Adding and using (\ref{minmi})
\begin{eqnarray}
\sum_{i=0}^{N-2}B_i &\geq& (Nr-r_{N-1}) + (Nr-r_{N-1}-r_{N-2}) + \cdots \nonumber \\
 &\geq& (Nr-B) + (Nr-2B) + \cdots \label{dhgso} \\
 &=& N(N-1)r - \alpha B \label{ygjlq}
\end{eqnarray}
From (\ref{fiejs}),
(\ref{sdidk}) and (\ref{minmi}), we cannot have $r_i = B$ $\forall$ $i$.
So, $\alpha$ cannot be of the
order of $N^2$.  Thus, the lower bound on $\sum_{i=0}^{N-2} B_i$ given by
(\ref{dhgso}) and (\ref{ygjlq}) is a loose lower bound.  From (\ref{ukhef}),
(\ref{sdidk}) and (\ref{ygjlq}), $\sum_{i=0}^{N-2} B_i$
grows as $N^2$ and is upper-bounded  by $5(N-1)r$, which is 
impossible, except when $N$ is  small. 
So, we discard Case 2.

From the result of Case 1, $H_0^{*'}(0) > H_0^*(0)$.  So, our
assumption that ${\mathcal R}$ is an optimal GTBR is
incorrect. Therefore, equality must hold in (\ref{ukhef}) for every
optimal GTBR.  \hfill $\blacksquare$

\section{Optimal GTBR}
\label{simulation1}

\begin{figure}[htbp]
{\centering \resizebox*{3.2in}{!}{\includegraphics{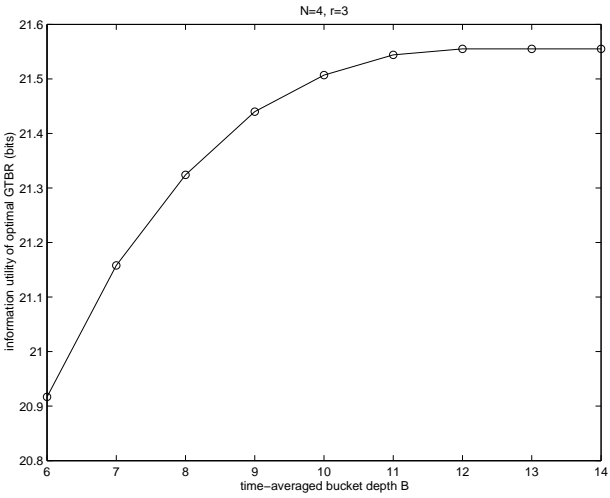}} \par}
\caption{$H_g^*$ vs. $B$ is concave.}
\label{hvsb}
\end{figure}

\begin{figure}[hbtp]
{\centering \resizebox*{3.2in}{!}{\includegraphics{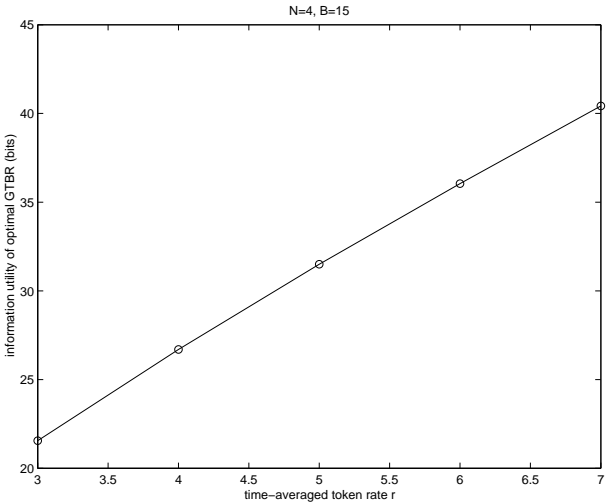}} \par}
\caption{$H_g^*$ vs. $r$ is highly linear.}
\label{hvsr}
\end{figure}

\begin{table}[hbtp]
\small
\begin{tabular}{|c|c|c|c|c|c|} \hline
($N$,$r$,$B$) & ${\mathbf r^*}$ & ${\mathbf B^*}$ & $H_s$ & $H_g^*$ & inc. \\
 &  &  & (bits) & (bits) & (\%) \\ \hline
(4,3,6) & ($\mbox{6 3 3 0}$) & ($\mbox{6 6 6}$) & 20.04  & 20.92  & 4.4 \\ \hline
(4,3,9) & ($\mbox{8 3 1 0}$) & ($\mbox{8 10 9}$) &   &   &  \\
 & ($\mbox{9 2 1 0}$) & ($\mbox{9 10 8}$) & 20.10  & 21.44  & 6.7 \\ \hline
(4,3,12) & ($\mbox{12 0 0 0}$) & ($\mbox{12 12 12}$) & 20.10  & 21.56   & 7.2 \\ \hline
(4,4,8) & ($\mbox{8 4 4 0}$) & ($\mbox{8 8 8}$) & 25.08  & 26.04  & 3.8 \\ \hline
(4,4,10) & ($\mbox{9 5 2 0}$) & ($\mbox{9 12 9}$) & 25.13  & 26.39  & 5.0 \\ \hline
(4,4,12) & ($\mbox{11 4 1 0}$) & ($\mbox{11 14 11}$) & 25.14  & 26.59  & 5.8 \\ \hline
(4,4,16) & ($\mbox{16 0 0 0}$) & ($\mbox{16 16 16}$) & 25.14  & 26.70  & 6.2 \\ \hline
(4,5,10) & ($\mbox{10 5 5 0}$) & ($\mbox{10 10 10}$) & 29.91  & 30.92  & 3.4 \\ \hline
(4,5,12) & ($\mbox{11 6 3 0}$) & ($\mbox{11 14 11}$) & 29.96  & 31.24  & 4.3 \\ \hline
(4,6,12) & ($\mbox{11 7 6 0}$) & ($\mbox{11 13 12}$) &  &  &  \\
 & ($\mbox{12 7 5 0}$) & ($\mbox{12 13 11}$) & 34.60  & 35.66  & 3.1 \\ \hline
(5,3,6) & ($\mbox{6 3 3 3 0}$) & ($\mbox{6 6 6 6}$) & 25.68  & 26.57  & 3.5 \\ \hline
(5,3,9) & ($\mbox{8 3 3 1 0}$) & ($\mbox{8 10 10 8}$) & 25.88  & 27.33  & 5.6 \\ \hline
(5,3,12) & ($\mbox{11 2 2 0 0}$) & ($\mbox{11 13 13 11}$) & 25.90  & 27.59  & 6.5 \\ \hline
(5,3,15) & ($\mbox{15 0 0 0 0}$) & ($\mbox{15 15 15 15}$) & 25.90  & 27.64  & 6.7 \\ \hline
(6,3,6) & ($\mbox{6 3 3 3 3 0}$) & ($\mbox{6 6 6 6 6}$) & 31.33  & 32.23  & 2.9 \\ \hline
\end{tabular}
\caption{Entropy-maximizing GTBR for given $N$, $r$ and $B$.}
\label{kdhfo}
\end{table}
\noindent We determined the optimal GTBR by exhaustive search over the
reduced search space obtained from the proposition. Our computation
results are shown in Table \ref{kdhfo}. $H_s$ and $H_g^*$ denote the
information utility of the STBR ${\mathcal R}_s(N,r,B)$ and the
optimal GTBR ${\mathcal R}_g(N,{\mathbf r}^*,{\mathbf B}^*)$
respectively. Based on our computations, we infer:
\begin{enumerate}

\item
A generalized token bucket regulator can achieve {\it higher}
information utility than a standard token bucket
regulator. The increase in information utility is significant
(up to 7.2\%), esp. for higher values of $B$.

\item 
The optimal bucket depth sequence
${\mathbf B}^*$ is uniform or near-uniform, i.e.,
the standard deviation is very small compared to the mean.

\item 
The optimal token increment sequence
${\mathbf r}^*$ is a decreasing sequence and is not uniform.

\item
For a fixed $N$ and $r$:
\begin{enumerate}

\item 
If $B=2r$, ${\mathbf B^*}$ is always uniform and ${\mathbf r^*}$ is
uniform except for the terminal values.

\item 
\label{dflke}
As $B$ increases from $2r$ to $\min(5,N)r$, the variance of ${\mathbf
r^*}$ increases rapidly with a concentration of tokens in first few
stages, the variance of ${\mathbf B^*}$ increases slowly, while
$H_g^*$ initially increases and then saturates at some final value.
$H_g^*$ is an increasing and concave sequence\footnote{The first-order
differences form a decreasing, non-negative sequence.} in $B$ (Figure
\ref{hvsb}).

\end{enumerate}

\item 
\label{yrtyi}
For a fixed $N$ and $B$, $H_g^*$ is an increasing, highly linear
and slightly concave sequence in $r$ (Figure 
\ref{hvsr}). For the STBR, Results \ref{dflke} and \ref{yrtyi} 
have  been observed in \cite{iee_premal}.

\end{enumerate}

\section{Information-theoretic interpretation}
\label{mnbvc}

From classical information theory, if $\sum_{i=1}^n p_i =1$, system
entropy $H$ increases with decreasing Kullback-Leibler distance
between the given probability mass function (pmf) and the uniform
pmf. $H$ is maximized only if $p_1=\cdots=p_n=\frac{1}{n}$. Also,
maximum system entropy
$H^*$ increases with $n$ \cite{it_cover_thomas}.  Analogously, a GTBR
can achieve higher information utility than an STBR because the pmfs
of the packet lengths at each stage are more uniform and have a larger
support.  For a given $\mathbf r$ and $\mathbf B$, recall that
information utility is computed recursively by (\ref{tok-evlv}) and
(\ref{hkopti}).

We argue that ${\mathbf B^*}$ must be uniform or near-uniform for
maximum information utility. If ${\mathbf B^*}$ is neither uniform nor
near uniform, then $B_j = \min_i B_i$ is much smaller than $B$. This
restricts the range of values taken by $u_{j+1}$ and $\ell_{j+1}$
(from (\ref{tok-ineq}) and (\ref{tok-evlv})). The support of packet
length pmfs at stage $j+1$ is reduced, leading to lower flow entropy
at stage $j+1$ and consequently lower information utility.  Thus,
${\mathbf B^*}$ must be uniform or near-uniform to maximize the
minimum support of the packet length pmfs {\it at each stage}.  Also,
in Table \ref{kdhfo}, observe that $\min_{i}B_i^* = B-1$ or
$\min_{i}B_i^* = B$ throughout.

We now argue that for maximum information utility, ${\mathbf r^*}$
must be a decreasing sequence, subject to $r_i \leq B_i$ for every
$i$.  If $r_i > B_i$ for any $i$, then a zero length packet cannot be
transmitted in slot $i$ (from (\ref{tok-evlv})) and will have zero
probability. This decreases the support of the packet length pmfs in
slot $i$ and leads to lower information utility. Importantly, 
from (\ref{uwxphis})
\begin{eqnarray*}
H_0^*(0) &=& \sum_{\ell_0=0}^{r_0} p_{\ell_0}^*(0) \Big(
\ell_0 - \log_2(p_{\ell_0}^*(0)) + H_1^*(\min(r_0\\&&-\ell_0,B_0)) \Big)
\end{eqnarray*}
The major contribution to information utility $H_0^*(0)$ is from the
support of the packet lengths $[0,r_0]$ and the pmf of the packet
lengths ${\mathbf p}_0^*(0)$, while the contribution from
$H_1^*(\cdot)$ is insignificant.  So, to maximize $H_0^*(0)$, $r_0$
should be allowed to take its maximum possible value, subject to $r_0
\leq B_0$, and the pmf of the packet lengths should be close to the
uniform pmf.  The observation that $r_0=B_0$ consistently in Table
\ref{kdhfo} corroborates this.  Also, a high value of $r_0$ leads to
larger supports of packet length pmfs at intermediate and
later stages. Similarly, the
first few elements of $\mathbf r^*$ tend to take large values till the
aggregate tokens are exhausted.  However, their contribution to
$H_0^*(0)$ is not as pronounced and equality may not hold in $r_i \leq
B_i$. Thus, ${\mathbf r^*}$ must be a decreasing sequence and the
first few elements of ${\mathbf r^*}$ tend to take their maximum
possible values, subject to $r_i \leq B_i$, to achieve uniformity and
larger supports of packet length pmfs {\it at intermediate and later stages}.

This ``greedy'' nature of ${\mathbf r^*}$ is evident when $N$ and $r$
are kept constant and $B$ increases (Result \ref{dflke}). A similar
argument is applicable when $N$ and $B$ are kept constant and $r$
increases (Result \ref{yrtyi}).  The only difference is that a unit
increase in $r$ will necessarily increase $H_g^*$ by at least $N$ bits
($N$ bits are contributed by the packet contents alone, which also
explains the dominant linear variation in Figure \ref{hvsr}), while a
unit increase in $B$ will increase $H_g^*$ only by an amount equal to
the difference in covert information.  The increase in covert
information is positive only if the resulting optimal token increment
and bucket depth sequences $({\mathbf r^*},{\mathbf B^*})$ result in
larger support and more uniformity for the packet length pmfs.
Indeed, when $B$ increases beyond the maximum number of tokens
possible at any stage ($\max_i\{\phi_i\}$), clamping the residual
number of tokens at every stage becomes ineffective and the system
behaves as if bucket depth constraints were not imposed at all (Figure
\ref{hvsb}).

\section{Discussion}
\label{dksjl}
In this paper, we have considered a problem where a source whose
traffic is regulated by a generalized token bucket regulator, seeks to
maximize the entropy of the resulting flow.  The source can achieve
this by recognizing that the randomness in packet lengths acts as a
covert channel in the network and sizing its packets appropriately. We
have formulated the problem of computing the GTBR with maximum
information utility in terms of constrained token increment and bucket
depth sequences. A GTBR can achieve higher information utility than a
standard IETF token bucket regulator.  Finally, we have
information-theoretically interpreted the observation that an
entropy-maximizing GTBR always has a near-uniform bucket depth
sequence and a decreasing token increment sequence.

Our results show the existence of upper bounds on the entropy of
regulated flows. It would be interesting to construct source codes
which come close to this bound.  The development of a rate-distortion
framework for a generalized token bucket regulator is 
currently under investigation.

\bibliography{gtbr}

\end{document}